# Life's Solutions are Not Ideal

Bob Eisenberg
Mathematics and Computer Sciences Division
Argonne National Laboratory
9700 South Cass Avenue
Argonne, IL 60439

Department of Molecular Biophysics and Physiology
Rush University
1653 West Congress Parkway
Chicago IL 60612

May 1, 2011

**Abstract**

Life occurs in ionic solutions, not pure water. The ionic mixtures of these solutions are very different from water and have dramatic effects on the cells and molecules of biological systems, yet theories and simulations cannot calculate their properties. I suggest the reason is that existing theories stem from the classical theory of ideal or simple gases in which (to a first approximation) atoms do not interact. Even the law of mass action describes reactants as if they were ideal. I propose that theories of ionic solutions should start with the theory of complex fluids because that theory is designed to deal with interactions from the beginning.

The variational theory of complex fluids is particularly well suited to describe mixtures like the solutions in and outside biological cells. When a component or force is added to a solution, the theory derives — by mathematics alone — a set of partial differential equations that captures the resulting interactions self-consistently. Such a theory has been implemented and shown to be computable in biologically relevant systems but it has not yet been thoroughly tested in equilibrium or flow.

Ions in water are life's solutions.[48,49] Pure water is lethal to most cells. Most proteins denature in distilled water. Electrolyte mixtures in water—not water itself[30,38]—are the liquid of life.

The concentration and type of ions are important in biology.[11,19,52,102-104] Most cells live in a small range of ions and concentrations. Changing the type or concentration of ions changes the function of most enzymes, binding proteins, or channels, as well as cells and tissues. Details of the mixture of ions in our plasma and around our cells have profound effects. Just ask a clinician about blood *pH*.

Surprisingly little is known about the physical properties of salt mixtures like those inside and outside cells. A prominent chemist recently reported (in an overview of specific ion effects[80]) "it is still a fact that over the last decades, it was easier to fly to the moon than to describe the free energy of even the simplest salt solutions beyond a concentration of 0.1M or so." Experimental measurements have been available for some 100 years. A generation of physical chemists sought explanations but only produced empirical equations for a subset of phenomena[1,2,4,6,7,14,15,22-25,34-37,39,40,42,44,46,61,63,66-69,71-80,82,83,85-87,91-96,99,100,106,107,109-112,114,115]. Realistic simulations are just beginning to address the issues in the simplest (unrealistic) cases[2,36,37,42,47,61,62,67,71,72,76,79,80,93,100,107,112,115], rarely dealing with divalents, mixtures, or flow.[15,70,100] Many physical chemists have told me of their frustration at this lack of understanding. I am alarmed as a biologist that life's solutions are so little understood.

**Theories of the Ideal**. Classical theories of ionic solutions are of limited use because they stem from the theory of ideal gases. In one of the triumphs of 19$^{th}$ century physics, physicists showed how gases could be approximated as swarms of point particles without electrical properties,[9,10,13,45,101] extending that triumph to the liquid state.[5,45,97] Chemists and biologists naturally tried to extend the triumph of ideal gas theory to ions in water [2,36,37,42,47,61,62,67,71,72,76,79,80,93,100,107,112,115] despite the evident difference in density, charge, and thus interactions. That extension is used to this day.

But life's solutions are nothing like these ideal gases of noninteracting particles. All electrolyte solutions—even those as dilute and nearly ideal as 1 mM $Na^+Cl^-$—are dominated by the electrical interactions called shielding or screening.[7,25,28,34,36,45,46,61,69,79,83,95,110] Their free energy does not scale linearly with the number density of particles as in ideal gases. The free energy contains a large 'excess' component that (for $Na^+Cl^-$) scales with the square root of concentration. Even in dilute solutions 'everything' interacts with everything else through the mean electric field.

**Theories of the Dilute**. Dilute (< 150 mM) solutions of $Na^+Cl^-$ and their mean electric fields are usually described by the Poisson Boltzmann equation. The PB equation idealizes ions as point charges. It appears in various disguises as the Gouy-Chapman equation of planar solution interfaces and, linearized in spherical coordinates, as the Debye-Hückel theory of dilute solutions of spherical ions. PB equations only describe equilibrium conditions in which nothing flows.

Extensions of PB to include flow have many names. Drift diffusion equations are used to design our semiconductor technology.[64,88,105] The Poisson-Nernst-Planck equations of physical chemistry and biophysics[8,16,20,32,33] are drift diffusion equations often called PNP[31] to emphasize the relation[26] between semiconductors and ionic solutions.

The PNP treatment is seriously inadequate because ions are not points. Ions cannot overlap. The finite diameter of ions has a surprisingly large effect on electrostatic energies because the energy of the electric field of a swarm of spheres is not well approximated by the electric field of points. The electric field of a sphere deviates from that of a point just where the field is strongest, near the origin. An arbitrary distance of closest approach has to be added to make PB or PNP at all reasonable.

A generation of physical chemists just now retiring studied these effects, reported in the classics of the statistical mechanics literature [5,45,97]. The scientists may retire, but finite size effects have not disappeared. As one of that generation — George Stell — put it "…it is almost never valid to use Debye-Hückel under physiological conditions" (paraphrase of [110]). This is a view repeated many time by many workers of that generation, see [7,25,28,34,36,47,79,83,95] and other references too numerous to cite here. Two large books[76,114] compile the relevant experimental data for equilibrium systems and document the many attempts to fit it along with references already cited.

The idealizations fail because the finite size of one ion type changes the free energy of other types. **No ion type can be treated independently of the others.** It is not correct to write equations in which a property of one ion depends on the concentration of just that ion. The electrochemical potential of one ion depends on all other ions. The independence principle of Hodgkin and Huxley is incorrect in bulk solutions and inside channels that are occupied by more than one type of ion. [52-60] Hundreds of papers reviewed in [7,25,28,34,36,47,76,79,83,95,114] (to choose only a few citations) show that the concentration of one ion changes properties of other types of ions. Everything interacts with everything else. All the ions interact with each other through the electric field, and through crowding. Mixtures of ions are nothing like ideal.

Little trace of these facts appears in the conventional wisdom of molecular biology and biophysics, or their texts[3,18,50], probably because mathematical descriptions of interactions were so arbitrary and awkward in traditional theories.[114]

**Interactions in Crowded Charged Solutions**. Interactions should be described naturally and self-consistently in mathematical models of real solutions. Traditional models deal with interactions in an *ad hoc* and often subjective way. They add interactions one at a time and use a large number of ill-determined coefficients that change as any component of the system is changed.[70,76,85,90,95,114] This paper proposes a mathematical approach — from the theory of complex fluids [21,81,84,113] — that objectively and automatically captures all interactions. This approach derives differential equations that describe all the forces, fields and flows (and interactions) with a minimal number of parameters.

Life's (ionic) solutions seem not to have been treated previously as complex fluids. The mathematics of this approach has been proven (in theorems and existence proofs) and has been shown to be practical and useful in physical applications. But the application to life's solutions appears to be new. I now argue that a variational approach is more or less essential when flows are involved or number densities of ions — that we call 'concentrations' — are high, in the molar region.

Ions are present at very high concentrations in most devices — physical or biological — that use them. Ions are packed closely next to electrodes in practical electrochemical devices. Ions are at enormous densities in ion channels because the fixed charge of acidic and basic side chains of proteins demands equal amounts of charge (and usually ions) nearby. (Small violations of even local electroneutrality would produce electric fields known to destroy ion channels and proteins.) Calcium and sodium channels have 20 to 40 molar concentrations of ions. For comparison, solid NaCl has a number density (in chemical units) of 37 molar. Active sites of enzymes have densities of acidic or basic side chains as large as ion channels. The charge density of side chains in active sites is ~20 molar in ~600 enzymes, found in an automated search of a library of enzymes of known structure and function.[65]

The crowded conditions of active sites and channels form a special environment likely to be important for biological function. The mathematics used to describe this crowded environment must deal naturally with interactions because everything is coupled. Everything interacts with everything else. Nothing is ideal. The free energy of one ion depends on the concentration of others.

**Law of Mass Action**. Even the law of mass action — taught so early in our educations that many take its validity for granted — depends on the ideal properties of reactants.[95] The derivation of the law uses expressions for energy and entropy of ideal gases of uncharged point particles. Indeed, the rate constants of the law of mass action can be constants, independent of concentration, **only** if the reactants are ideal, dilute, and their free energy is described by simple formulas that depend only on the number density (and charge) of one species at a time. Special conditions can be chosen to approximate these ideal conditions, and these conditions are widely used in idealized measurements of enzyme kinetics, but channels and enzymes rarely actually function in these special conditions. Channels and enzymes work in mixtures of ions, with tightly specified concentrations of 'control' molecules often including calciuim ions.

When reactants interact, the free energy of one type of reactant depends on the concentrations (and types) of all species present.[28,76,85,90,114] The free energy then has a component in excess of that in an ideal gas. **The excess component is important whenever reactants are charged.** It dominates when reactants are at high number density as they are in the immediate vicinity of catalysts, or enzymes, or if they have a 'high cross section for reaction' for any reason. Indeed, as chemical reactions are studied in three dimensions, without assuming well stirred "zero dimensional" solutions, it will be interesting to see how many reactions occur at low concentration. One imagines that most chemical reactions that proceed at a reasonable and interesting rate have local concentrations of reactants near enzymes or catalysts (or near other reactants) that are large. The concentrations of reactants in these interesting regions are unlikely to be those assumed in approximations that assuming perfect stirring, in my view. Engineers and evolution are likely to choose systems that maximize the rate of reactions.

The law of mass action will have variable reaction rate 'constants' whenever reactants are charged. It will have variable reaction rate 'constants' whenever a reaction proceeds rapidly because of a large local concentration of reactants.

We conclude that life's reactions, likes its solutions, often need to be analyzed by a mathematics that deals naturally with interactions. Fortunately, the (mathematical variational) theory of complex fluids[84,113] does this. The theory has been developed by mathematicians interested in abstract as well as practical issues: how to derive Navier Stokes equations from variational principles. The theory has been used by physicists interested in liquid crystals and other systems containing complex components that change shape and size, even fission and fuse, as they flow in wildly nonequilibrium systems. The flows successfully computed by the theory of complex fluids can be much more complicated than those we see every day when water flows from a faucet. The liquid crystal displays (LCD's) of our technology are computed by these theories.[21,81]

**Electrolyte Solutions as Complex Fluids**. I suggest that work on electrolyte solutions should start with the theory of complex fluids[21,81,84,113], beginning with the relatively simple case of

*LifeSolArXiv3.docx* 3

mixtures of hard spherical ions, because that mathematics deals consistently and automatically with interactions. More complex representations of ions and solvent can be added later on.

The theory of complex fluids brings along another advantage. It can deal with nonequilibrium properties of ionic solutions. It deals with flows driven by pressure, concentration, and voltage gradients. Describing such flows has been the despair of physical chemists. The ionic interactions that cannot be described at rest are even harder to describe in flow.[15,70,100] Each flowing species has its own mobility, and each mobility depends on the concentrations, flows, and nonideal properties of all the other ions. Every nonequilibrium property interacts with every other nonequilibrium property, as well as every equilibrium property interacting with every other equilibrium property.

Dealing with nonequilibrium systems is particularly important in life. Nearly every system in biology requires nonequilibrium conditions and flows of ionic solutions, as does manmade technology. Living systems live only when they maintain large flows to keep them far from equilibrium. Electron flows are the electrical signals in coaxial cables of our technology. Ionic flows are the electrical signals in the cables (axons) of our nervous system. Our technology is like our life. Flows are needed to create the conditions that allow devices or axons to have simple properties. Amplifiers follow simple laws only when energized by power supplies. An amplifier with power supplies set to ground potential follows no simple law. Only when is it is energized is the output proportional to the input. It takes many highly nonlinear devices (usually transistors) working far from equilibrium to produce that simple relation between input and output. In engineering, thermodynamic equilibrium only occurs when power supplies are turned off and devices no longer function. In biology, thermodynamic equilibrium only occurs at death. Theories developed to exploit the special properties of thermodynamic equilibrium had an enormously important role in the historical development of science,[12,41] but they must be abandoned in their original form, if the interacting nonequilibrium reality of life's solutions are to be explained as they are actually measured. One can argue that semiconductor technology was possible exactly because semiconductor physicists never assumed equilibrium [64,98,108] (Compare the constant field theory in semiconductors[89] and with the physiological version[11,19,52,102-104] by Hodgkin and Katz, and Cole's student Goldman.[17,43,51]

**Flow**. The variational theory of complex fluids is likely to help in the analysis of flow. Variational principles can describe the energy and friction ('dissipation') of the components of complex fluids. The field equations derived from these principles describe spatial variation and flow within a set of boundaries with known properties, like the electrodes or the insulating walls of an electrochemical cell. The field equations are not restricted to chemical equilibrium. They deal with flow as they deal with no flow. Both are consequences of the constituents as expressed in the underlying variational principle and the boundary conditions. The partial differential equations that form a field theory are derived from variational principles by mathematics alone. No approximations or additional arguments are needed to describe interactions.

If a new component is added to a complex fluid, the component interacts with everything else in the system, and the resulting analysis must reflect that interaction. These interactions have rarely been guessed in full when analysis begins with the differential equations instead of a variational principle. There are too many possible descriptions and too many interactions that are easily overlooked.

The interactions arise automatically from a variational principle after the Euler-Lagrange process is used to derive partial differential equations of field theory. The interaction terms are objective outputs of a mathematical analysis. They are not assumed. If a new component is added into the variational principle, the interactions of the components arise automatically in the resulting partial differential equations.

Of course, the variational principle is not magic. If the components of the energy or dissipation are incorrectly described, the field equations will be incorrect, albeit self-consistent mathematically. For example, if a chaotropic additive changes the free energy of water, a primitive model[7,34,36,83] of an ionic solution (with constant free energy of water) will fail. We cannot know how important such effects are from mathematics. A physical theory must be constructed and compared to specific experimental data to see if it actually works. The field equations of a variational theory will always be consistent mathematically, but they will be consistent physically only if they contain enough detail. We do not yet know what detail is needed in the solutions of life.

This variational approach is in contrast to attempts to simulate macroscopic properties by direct calculation of motions of all atoms in molecular dynamics. The problems of scale are much larger and more serious than often discussed, see reference[27]. The problems of dealing with life's structures (with 0.1% accuracy in three dimensions) that change significantly in $10^{-14}$ sec are serious, particularly because the structures perform their natural functions by moving on the $10^{-4}$ sec time scale. In quantitative applications, molecular dynamics is limited (for the time being) by its inability to calculate the properties of mixtures of electrolytes. In qualitative applications, molecular dynamics provides an indispensable dynamic extension of the statics of structural biology.

**Variational Analysis**. The variational analysis of ionic solutions, and ions in channels, can begin with the energetic variational principle of Chun Liu [84,113], building on the work of many others,

$$\overbrace{\frac{\delta E}{\delta \vec{x}}}^{\textit{Conservative 'Force'}} - \overbrace{\frac{1}{2}\frac{\delta \Delta}{\delta \vec{u}}}^{\textit{Dissipative 'Force'}} = 0 \qquad (1)$$

This principle has been applied to a model of ionic solutions in which the ions are represented as spheres in a uniform frictional dielectric.[29,62] This primitive model of an ionic solution — that uses a dielectric to describe implicitly the properties of water — is well-precedented and more successful than most in dealing with experimental data, but it will certainly need to be extended to include much more atomic detail.[61].

If ions are modeled as Lennard Jones spheres, the variational principle produces 'Euler Lagrange' equations of a drift-diffusion theory with finite sized solutes, a generalization and correction of PNP.

$$\frac{\partial c_n}{\partial t} = \nabla \cdot \left[ D_n \left\{ \nabla c_n + \frac{c_n}{k_B T} \left( z_n e \nabla \phi - \int \frac{12\varepsilon_{n,n}(a_n + a_n)^{12}(\vec{x}-\vec{y})}{|\vec{x}-\vec{y}|^{14}} c_n(\vec{y}) d\vec{y} \right. \right. \right.$$
$$\left. \left. \left. - \int \frac{6\varepsilon_{n,p}(a_n + a_p)^{12}(\vec{x}-\vec{y})}{|\vec{x}-\vec{y}|^{14}} c_p(\vec{y}) d\vec{y} \right) \right\} \right] \qquad (2)$$

combined with the Poisson Equation

$$\nabla \cdot (\varepsilon \nabla \phi) = -\left(\rho_0 + \sum_{i=1}^{N} z_i e c_i\right) \qquad i = n \text{ or } p \qquad (3)$$

We write the equation only for negative monovalent ions with valence $z_n = -1$ to keep the formulas reasonably compact. Programs have been written for all valences. $c_{p,n}(\vec{y})$ is the number density of positive $p$ or negative $n$ ions at location $\vec{y}$. $D_{p,n}$ is the corresponding diffusion coefficient. $k_B T$ is the thermal energy, $k_B$ the Boltzmann constant, and $T$ the absolute temperature. $\varepsilon_{n,n}$ and $\varepsilon_{n,p}$ are coupling coefficients. $a_{p,n}$ are the radii of ions. $\varepsilon$ (without subscript) is the dielectric coefficient. The electrical potential is $\phi$. $\rho_0$ represents the charge density of the protein, as it represents the charge density of doping in semiconductors. $\rho_0$ depends on location and is zero in bulk solutions.

We note that the field satisfies the dissipation principle.

$$\overbrace{\frac{d}{dt}\int\left\{k_B T \sum_{i=n,p} c_i \log c_i + \tfrac{1}{2}\left(\rho_0 + \sum_{i=n,p} z_i e c_i\right)\phi + \sum_{i,j=n,p}\frac{c_i}{2}\int \tilde{\Psi}_{i,j} c_j d\vec{y}\right\}d\vec{x}}^{\text{Dissipative}}$$

$$= -\underbrace{\int\left\{\sum_{i=n,p}\frac{D_i c_i}{k_B T}\left|k_B T \frac{\nabla c_i}{c_i} + z_i e \nabla \phi - \sum_{j=n,p}\nabla\int \tilde{\Psi}_{i,j} c_j d\vec{y}\right|^2\right\}d\vec{x}}_{\text{Conservative}} \qquad (4)$$

$\tilde{\Psi}_{i,j}$ represent the Lennard Jones crowded charges terms defined in [29,62].

These equations have been integrated numerically in the papers cited to predict binding of ions in crowded conditions, or the flow of ions through channels. The calculations are successful and the methods are feasible. But numerical problems limit the routine application to realistic structures and so optimization of parameters are not yet possible. Until parameters are optimized one unfortunately cannot tell how well the theory actually deals with data.

**Conclusion**. The variational treatment of ionic solutions has substantial advantages. It describes systems in which everything interacts with everything else, as they do in life and its solutions. The variational theory can be a complete description of all the properties of ionic solutions in flow and in equilibrium. It can be a good approximation to the properties of ions in and around proteins when the protein has a well defined average structure.

The variational treatment of electrolyte solutions has just started. It begins in the right place, in my view, with the successful theory of complex fluids. But the variational approach is not yet sufficiently developed or checked. Other approaches are being used[2,36,37,42,47,61,62,67,71,72,76,79,80,93,100,107,112,115] (and many overlooked or forgotten by me, no doubt)

to deal with solutions of a single monovalent species at equilibrium and one hopes for their early success. It is not clear, however, that other approaches can deal with divalents like $Ca^{2+}$, or with multicomponent systems with ions of unequal diameter and charge like plasma or blood, or with flow at all. These properties are essential to life. Analysis must not assume properties incompatible with life: life's solutions are flowing mixtures of unequal diameter and variational methods deals naturally with these complexities and interactions.

      The ionic mixtures of life's solutions need to be understood. Life depends on the bulk flow and electrodiffusion of multicomponent systems. It is often controlled by trace concentrations of messengers, often divalent calcium ions. We need to calculate the properties of sea water, or the plasmas in our body, with a physical theory that lets us understand and change what these solutions do to our proteins, cells and tissues. Variational methods give hope, at least to me, that this can be done.